\documentclass[format=acmsmall, review=false, screen=true]{acmart}
\usepackage[utf8]{inputenc}
\usepackage{booktabs} 

\usepackage[ruled]{algorithm2e} 
\usepackage{booktabs}

\usepackage{courier}
\usepackage{url}
\usepackage{graphicx}
\usepackage{color}
\usepackage{array}
\usepackage{subfigure}
\usepackage[T1]{fontenc}
\usepackage{aecompl}

\SetAlFnt{\small}
\SetAlCapFnt{\small}
\SetAlCapNameFnt{\small}
\SetAlCapHSkip{0pt}
\IncMargin{-\parindent}

\newtheorem{resques}{Research Question}

\received{January 2020}
\received[revised]{June 2020}
\received[accepted]{July 2020}

\begin{document}
\title[Understanding the Role of Intermediaries in Online Social E-commerce:\\ An Exploratory Study of Beidian]{Understanding the Role of Intermediaries in Online Social E-commerce: An Exploratory Study of Beidian}


\author{Zhilong Chen}
\affiliation{
  \institution{Beijing National Research Center for Information Science and Technology (BNRist), Department of Electronic Engineering, Tsinghua University}
  \city{Beijing}
  \country{China}}
  
\author{Hancheng Cao}
\affiliation{
  \institution{Department of Computer Science, Stanford University}
  \city{California}
  \country{United States}}
 
\author{Fengli Xu}
\affiliation{
  \institution{Beijing National Research Center for Information Science and Technology (BNRist), Department of Electronic Engineering, Tsinghua University}
  \city{Beijing}
  \country{China}}
  
\author{Mengjie Cheng}
\affiliation{
  \institution{Harvard Business School}
  \city{Massachusetts}
  \country{United States}}
  
\author{Tao Wang}
\affiliation{
  \institution{Grenoble Ecole de Management}
  \city{Grenoble}
  \country{France}}
 
\author{Yong Li}
\email{liyong07@tsinghua.edu.cn}
\authornote{Corresponding author.}
\affiliation{
  \institution{Beijing National Research Center for Information Science and Technology (BNRist), Department of Electronic Engineering, Tsinghua University}
  \city{Beijing}
  \country{China}}

\newcommand{\hancheng}[1]{{\textcolor{red}{ [hancheng: #1]}}}
\setcopyright{none}
\begin{abstract}
Social e-commerce, as a new form of social computing based marketing platforms, utilizes existing real-world social relationships for promotions and sales of products. It has been growing rapidly in recent years and attracted tens of millions of users in China. A key group of actors who enable market transactions on these platforms are intermediaries who connect producers with consumers by sharing information with and recommending products to their real-world social contacts. Despite their crucial role, the nature and behavior of these intermediaries on these social e-commerce platforms has not been systematically analyzed. Here we address this knowledge gap through a mixed method study. Leveraging 9 months' all-round behavior of about 40 million users on Beidian -- one of the largest social e-commerce sites in China, alongside with qualitative evidence from online forums and interviews, we examine characteristics of intermediaries, identify their behavioral patterns and uncover strategies and mechanisms that make successful intermediaries. We demonstrate that intermediaries on social e-commerce sites act as local trend detectors and ``social grocers''. Furthermore, successful intermediaries are highly dedicated whenever best sellers appear and broaden items for promotion. To the best of our knowledge, this paper presents the first large-scale analysis on the emerging role of intermediaries in social e-commerce platforms, which provides potential insights for the design and management of social computing marketing platforms.

\end{abstract}

\ccsdesc[500]{Human-centered computing~Empirical studies in collaborative and social computing}

%
%

\keywords{Behavioral analysis; Social e-commerce; Intermediary; Role}


\maketitle

\renewcommand{\shortauthors}{Zhilong Chen et al.}

\section{Introduction}

The conundrum of leveraging social relationships for online purchases in e-commerce has received great attention from both the industry and the academia in the past decade ~\cite{king2010introduction,liang2011introduction,liang2011drives,huang2013commerce}. While some research resorts to implicitly incorporating social relationships into recommendation systems and develops a branch of social recommendation~\cite{king2010introduction}, other e-commerce actors explicitly take social relationships into the original design of platforms, enabling the rise of social commerce~\cite{zhou2013social,busalim2016understanding}.
These social commerce platforms take two main approaches: adding social features onto their sites to unleash the power of Web 2.0 design features and tools~\cite{huang2013commerce}, or creating virtual customer environments more directly by building social media communities and fostering interactions with and among customers~\cite{culnan2010large}. 

\begin{figure} [t]
\centering
\subfigure{\includegraphics[width=.49\textwidth]{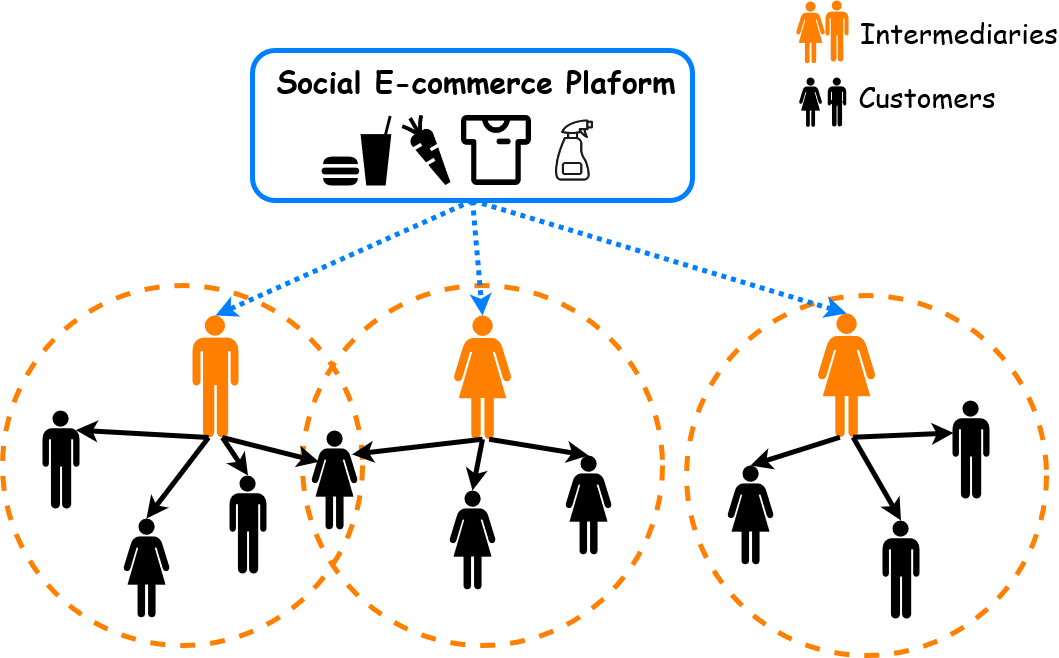}}
\caption{Illustration of intermediaries, and their relationships with social e-commerce platforms and customers. The solid black arrows denote recommendations/promotions which may eventually turn to transactions and the dotted blue arrows denote intermediaries' selecting items from social e-commerce platforms for recommendation and promotion. The dashed orange circles represent social cycles of intermediaries where they promote commodities.} \label{Mechanism}
\end{figure}

As social elements are increasingly incorporated in e-commerce, an emerging type of \emph{social e-commerce} 
is gaining momentum among different formats of social commerce. Extending the focus on being social, these platforms proactively engage in mobilizing the general public in promoting products and sharing information via their use of mobile devices. In particular, a new type of WeChat-based social e-commerce penetrates and leverages existing real-world close social relationships in people's instant messaging apps to reach family members, relatives, friends and colleagues. Consequently, people who actively share such information become market \emph{intermediaries} between platforms and potential customers. Fig.~\ref{Mechanism} shows the relationships across social e-commerce platforms, intermediaries and customers. These intermediaries share item links and recommend items on instant messaging platforms, bridging the supply end of social e-commerce platforms and the demand end of purchasers. These social e-commerce platforms have achieved great success in a relatively short period such as Pinduoduo\footnote{https://www.pinduoduo.com/}, Yunji\footnote{https://www.yunjiweidian.com/}, and Beidian\footnote{https://www.beidian.com/}, where the critical role of intermediaries makes these social e-commerce sites distinctive from the traditional ones and is essential in shaping these social e-commerce platforms' success. But who are these market intermediaries? How do their actions affect social e-commerce sites and market transactions? In this paper, we aim at uncovering the behaviors of these intermediaries and shedding light on their roles on social e-commerce platforms.

To better delineate the role of intermediaries, we investigate what strategies and forms intermediaries use to bind social e-commerce platforms and users in an exploratory manner.   
We adopt a mixed-method approach through combining quantitative results from a large-scale dataset and qualitative evidence to guarantee interpretability and verifiability at the same time. Our quantitative studies are conducted with a full-scale user information dataset and a 9-month user behavior dataset from Beidian, a leading social e-commerce platform in China. Complementary to quantitative investigations, we analyze 35 top posts from the section of Sales Skills of online intermediary forum Beidian Business School and interview 15 randomly-sampled Beidian intermediaries for qualitative studies. 

Specifically, we demonstrate that 
intermediaries act as local trend detectors: we spot positive correlations between consumers' past purchase preferences and intermediaries' recommendation preferences both on the level of a certain intermediary and on a community level. 
Qualitative evidence attributes the phenomena to 1) local need and interest detection and 2) borrowing other intermediaries' success from their intermediary communities. 
In addition, we demonstrate that these intermediaries take the role of "social grocers". We examine the distribution of customer number and deal number per intermediary-consumer pair, gauge the distribution of sales volume and average per-deal price of consumer-intermediary pairs, show the distribution of the number of categories sold by intermediaries and reveal geographical similarities of intermediaries and their consumers. Quantitative results together with qualitative evidence show that in terms of customer number and deal number, sales volume, average value and item category of deals, intermediaries' actions resemble characteristics of convenience stores/grocery stores. However, their sales differ from convenience stores/grocery stores in that they break geographical constraints, where physical closeness is substituted by social closeness.

We further analyze the characteristics of successful intermediaries to better understand intermediaries' roles. We classify them into 8 clusters based on their behavioral characteristics of re-recommendation tendencies and analyze the purchase-related metrics (e.g. sales) across clusters to identify successful intermediaries quantitatively, and utilize experiences of intermediaries who are considered successful officially for qualitative explorations. We examine 
these intermediaries' recommendation strategies and sales characteristics, concluding that successful intermediaries are dedicated to and only to local best sellers and turn their recommendations to resemble socially-connected convenience store, \emph{i.e.}, social grocers, better than their peers. 

To the best of our knowledge, our work provides the first large-scale analysis of the emerging role of intermediaries on social e-commerce platforms. Through studying what roles intermediaries are taking and what roles should be taken so as to make the best of their role, our findings contribute to better modeling of intermediaries and understanding of the success of social e-commerce, which can potentially motivate better design of computer-supported cooperative work (CSCW) and social computing based marketing platforms.

\section{Related work}

\subsection{Social Commerce}
Social commerce typically refers to e-commerce that uses social media technology to help e-commerce transactions and activities~\cite{liang2011introduction,zhou2013social,curty2011social}. However, the concrete implementation of social commerce can significantly vary from one another. Some social commerce researchers relate to incorporating social media features in e-commerce sites. This includes Amazon customer reviews and Groupon~\cite{hughes2012growth}, which are sometimes termed as social shopping~\cite{stephen2010deriving}. Other researchers relate to the utilization of existing social networks on social media for sales, which are sometimes termed as social e-commerce. Facebook-based F-commerce and Twitter-based T-commerce are two representative instances~\cite{culnan2010large}. 

There has been abundant research on social commerce from both the academia and the industry~\cite{busalim2016understanding}. Researchers utilize Theory of Planned Behavior~\cite{shin2013user} and augmented technology acceptance model~\cite{shen2012social} to examine social commerce adoption, and analyze the relationships between culture and trust~\cite{ng2013intention}, social support and uncertainty~\cite{bai2015effect}, institution-based trust~\cite{lu2016examining}, social climate of friendship groups~\cite{sun2016does}, social presence~\cite{lu2016social}, social desire~\cite{ko2018social}, social commerce constructs~\cite{hajli2015social}, technological features~\cite{zhang2014motivates} and design~\cite{huang2017effects} and people's purchase intentions. Other social commerce researchers dedicate to uncover how some key factors (including reputation, size, information quality, etc.)~\cite{kim2013effects}, social commerce constructs and social support constructs~\cite{shanmugam2016applications} and prior experience~\cite{shi2015trust} contribute to users' trust, and how various factors affect information disclosure~\cite{sharma2014disclosing} and information sharing~\cite{liu2016empirical} in social commerce. 

\subsection{Market Intermediaries}
Market intermediaries, situated between supply and demand, are important actors in establishing, maintaining and regulating market dynamics for social e-commerce. As an integral connecting device, market intermediaries involve selling, advising, matching and evaluating~\cite{bessy2013power}. Previous studies have shown the existence of different types of market intermediaries~\cite{dukes2016online}. For example, e-commerce sites and platforms are market intermediaries that connect producers and consumers~\cite{chakravarty2014customer}. Online consumers' reviews~\cite{chen2008online} and other types of user-generated contents serve the function of market intermediaries in marketing communication. Market intermediaries are engaged in activities that help consumers in searching and obtaining information, and evaluating and comparing products on the one hand~\cite{chen2002referral}. On the other hand, they also help producers in gaining market intelligence, estimating demands, and adjusting products to serve different customer bases effectively~\cite{pazgal2008behavior}. As our marketplace is increasingly populated by different evaluation devices such as rankings, ratings, and reviews, it poses challenges for both consumers and producers to sort and select information~\cite{bonebrake2002college}. Consequently, the credibility and accountability of market intermediaries and their recommendations become critical and have received broad scholarly attention.

\subsection{Social Recommendation}
Social recommendation~\cite{king2010introduction} - the utilization of social relations to recommender systems, has been identified as a means to better unleash the power of social issues to benefit recommendation outcomes. Some researchers reach out for considering social relationships algorithmically, where the emphasis on "social" based on theories of social influence~\cite{marsden1993network} and homophily~\cite{mcpherson2001birds} has been proved to be beneficial to recommendation performances~\cite{tang2013social,golbeck2006generating}. Attempts have been made to improve the effectiveness of social recommendations, such as co-factorizing user-item rating and user-user social relations~\cite{ma2008sorec}, aggregating users' and their friends' ratings~\cite{ma2009learning}, utilizing social relationships for regularization~\cite{ma2011recommender,lin2018recommender}, incorporating the propagation of trust~\cite{jamali2010matrix}, mapping users to low-dimensional trust spaces~\cite{yang2016social}, considering the concept of strong ties and weak ties~\cite{wang2016social} and tie preferences~\cite{wang2017learning}, and integrating social relationships in social media to cross-platform recommendations~\cite{lin2019cross}. Recently, some more advanced methods are utilized for improving social recommendations. Fan, Li and Zhang~\cite{fan2018deep} leverage deep neural networks to model social relationships and integrate the representation to rating matrix factorization. Wu et al.~\cite{wu2018collaborative} introduce collaborative neural social recommendation, model social relationships and user-item interactions through neural architecture altogether. Fan et al.~\cite{fan2019graph} further utilize graph neural network (GNN) framework to model social recommendation, while Wang, Zhu and Liu~\cite{wang2019social} propose optimal limited attention mechanism for achieving better social recommendations. 

Different from these works, our paper intends to uncover the power of social recommendations where the recommenders turn from algorithms to real people through better understanding the role of market intermediaries. What's more, in contrast to existing studies that adopt a broad-brush approach in articulating what "social" means, our paper examines the case in which the "social" nature of intermediaries is much narrower defined. 
Rather than virtual consumer communities and their opinions, how close social contacts in real-world life including family members, relatives, friends and colleagues use social media tools and serve as market intermediaries is taken into consideration. As such, social recommendation practices may exhibit their distinctiveness from previous studies.  

Specifically, as the use of mobile devices has pervaded the social life in China, the emergence and growth of social e-commerce sites that leverage existing real-world social contacts to attract new customers reflects both the scarcity and the need for trustworthy market intermediaries~\cite{subramanian2016leveraging}. The incorporation of Chinese \emph{guanxi}~\cite{yang2011virtual} into market transactions provides a unique and exciting avenue to investigate the characteristics and behavior patterns of market intermediaries and uncover their outreach and impact on social e-commerce and thus social commerce.

Therefore, in this paper, we focus on a special kind of social e-commerce practice in this track, WeChat-based social e-commerce. In the Chinese context, compared to the general cases of social e-commerce, transactions on this emerging form of social e-commerce primarily rest on the use of mobile devices and social relationships are manifested deeply in social media. Firstly, selling and buying of this emerging form of social e-commerce is mostly conducted on WeChat, the most popular instant messaging application in China. Contacts on WeChat are mostly mirroring real-world relationships, especially close relationships~\cite{Cornell2016,wang2015dwelling,church2013s}, providing a higher degree of social embeddedness with strong and frequent ties. Secondly, the network structure of these social e-commerce customer base is decentralized and flat, because these platforms rely on ordinary people's connections in every life rather than Key Opinion Leaders (KOL) in traditional social commerce~\cite{cao2019your}.

Some recent works have taken the first steps into modeling the basic characteristics of these social e-commerce platforms at scale. Cao et al.~\cite{cao2019your} investigate one of these social e-commerce platforms' decentralized network structure, its growth mechanism, and consumers' purchase behavior with special focus on proximity and loyalty. Their work demonstrates the uniqueness of social e-commerce networks compared to other social media networks and e-commerce sites. Xu et al.~\cite{xu2019think} adopt a data-driven approach to model people's purchase behaviors on such platforms. They propose and validate several mechanisms - matching, social enrichment, social proof and price sensitivity that increase purchase conversion. Xu et al.~\cite{xu2019relation} further propose Relation-aware Co-attentive Graph Convolutional Networks (RecoGCN) to model the heterogeneity in relationships of social e-commerce networks. Extending their interest in social e-commerce platforms, we draw on large-scale data demonstrations and detailed qualitative analysis and focus on the role of market intermediaries, who are close social contacts in one's daily life.


Based on existing literature on social commerce, market intermediaries and social recommendation, in this paper, we aim at addressing the following research questions,

\begin{resques}[RQ\ref{resques:intermediary}] \label{resques:intermediary}
 What do intermediaries do in the emerging form of social e-commerce?
\end{resques}

\begin{resques}[RQ\ref{resques:successful}] \label{resques:successful}
 Why do some intermediaries stand out in the crowd?
\end{resques}

\section{Method and Dataset}
This work aims to adopt a mixed-method approach to provide an interpretable and verifiable analysis of the role of intermediaries on social e-commerce. Specifically, we implement data-driven analysis at scale to provide quantitative evidence, while we further conduct interview and forum analysis to qualitatively dig into the role of intermediaries in depth. We carry out our analysis on Beidian, one of the largest and fastest-growing social e-commerce.

\subsection{Background: Mechanism of Beidian}

\begin{figure} [t]
\centering
\subfigure[New purchases made per day in 6 months] {\includegraphics[width=.49\textwidth]{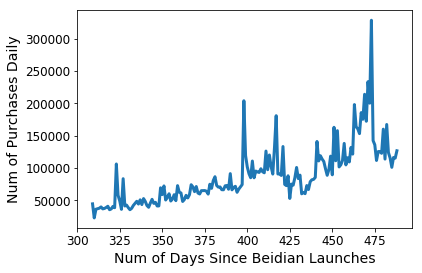}}
\subfigure[CDF of purchases made in 6 months] {\includegraphics[width=.49\textwidth]{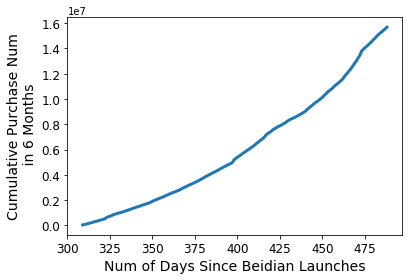}}
\caption{Total number of purchases on Beidian over 6 months' time.} \label{TimeGrowth}
\end{figure}

Beidian is one of the largest and fastest-growing social e-commerce\footnote{http://m.caijing.com.cn/api/show?contentid=4553034\&isappinstalled=0}. As depicted in Fig. \ref{TimeGrowth}, since its launch in August, 2017, Beidian has been growing at a high speed and its daily sales reached more than 320,000 after 1.5 years' expansion. More than 15,000,000 deals were completed on Beidian during 6 months, and the number of daily transactions on Beidian is continuously increasing.

One main characteristic that distinguishes Beidian is the unique business model which rests on matchmaking consumers and producers. More specifically, it mobilizes the general public through sharing product information with their real-world social contacts and facilitates communication and delivery of selected goods from producers to consumers\footnote{https://www.beidian.com/index.html}. On Beidian, people can assume two different yet overlapping roles after registration: consumers and intermediaries. The role of consumers denotes that people can buy goods and items on the platform. Intermediaries are individuals who actively promote Beidian's products through sharing/recommendation links on their instant messaging apps to their social contacts. Intermediaries are supported by Beidian to set up their own virtual shops and structure their recommendations, and act as intermediary agents between usual consumers and Beidian (and ultimately producers). Every user can become an intermediary after purchasing the designated goods that cost a small fee of roughly \$60.

Communication and transaction among consumers, intermediaries, and Beidian take place via WeChat, the most popular instant messaging app and social media platform in China. Intermediaries can share information and broadcast their recommendations through three channels: 1) WeChat Moments, which allow posting contents visible to one's friends; 2) WeChat groups, which enable communication within communities of no more than 500 members, and 3) WeChat messages of point-to-point contacts. Consumers click the recommendation and sharing links diffused by intermediaries. Once a deal is completed through an intermediary's link, he or she will receive an adequate portion of the deal from Beidian as remuneration. It is worth mentioning that without intermediaries' provision of links to Beidian, consumers cannot access products listed on Beidian. Therefore, intermediaries play a vital part in Beidian's access to a wider customer base and its eventual performance. Thus, understanding the characteristics and behaviors of these market intermediaries is critical in enriching our knowledge of the burgeoning social e-commerce phenomenon.

It is worth noting that many other top social e-commerce platforms adopt a similar mechanism, for example, Yunji\footnote{https://www.yunjiweidian.com/}, Miyuan\footnote{http://www.ppgps.cn/} and Fenxianglife\footnote{http://www.fenxianglife.com/}. Therefore, our results on Beidian have the potential capability of generalizing to other social e-commerce platforms of the kind, which contributes to a more comprehensive understanding of the whole social e-commerce ecology.

\subsection{Dataset for Quantitative Study}
We collaborate with social e-commerce platform Beidian to conduct our qualitative analysis, which enables our access to user information data and user behavior data which are utilized for quantifying intermediaries' behaviors.

\subsubsection{User Information Data}
The user information data covers the basic information of about 40 million users that have registered on Beidian before November 27th, 2018. Information including user's id, time of registration, type of identity (whether he or she has become an intermediary by November 27th, 2018) and person of invitation (who invites the user to Beidian) are provided. A total of 2.36 million users take the role of intermediaries, who account for a portion of 5.96\% of all users. For users who have made purchases on Beidian, their most frequently used cities and provinces for delivery are recorded.

\subsubsection{User Behavior Data}
The dataset of user behavior data incorporates overall user behaviors including clicking, cart-adding, purchasing and sharing/recommending. Records of purchasing and sharing span the period of May 31st, 2018 to March 6th, 2019. For purchase records, user ids of purchasers, user ids of corresponding intermediaries, items to be consumed and the time when the purchase deal was made are included. As a platform providing "home, clothing, food, cosmetics, maternal and child products and other good global goods for customers"\footnote{http://www.beidian.com/}, Beidian classify items into 12 categories: paper \& household cleaning, dietary supplements, baby clothes, snacks, fruits \& vegetables, milk powder \& diapers \& baby food, toys, cosmetics \& skincare, grains \& cooking oils \& drinks, household supplies, personal care and clothes \& shoes \& bags. More than 20 million purchase deals are completed during the period, which add up to a total sales volume of more than 800 million RMB yuan. About 180 million recommendations of items are made through the nine months' time.

\subsubsection{Ethical Considerations}
Careful steps are taken to address privacy issues concerning the sharing and mining of Beidian data which is utilized in this paper. Firstly, consent for research studies is included in the Terms of Service for Beidian. Secondly, preprocessing is done before we obtain the data, and user names and IDs are anonymized, through which user privacy is protected. Thirdly, our local university institutional board has reviewed and approved our research protocol. Finally, all data to be used is
stored in a secure off-line server. Access to the data is only limited to authorized members of the research team bound by strict non-disclosure agreements.

\subsection{Method and Data for Qualitative Study}
Qualitative studies are incorporated to strengthen our understanding of the role that intermediaries are playing on social e-commerce platform Beidian. Here we combine the data from the online forum Beidian Business School and an interview study on Beidian intermediaries for qualitative explorations.

Beidian Business School is an online forum integrated into Beidian application and is only accessible to Beidian intermediaries. The forum is divided into several sections, where we focus on the section of Sales Skills which discusses detailed selling skills that are perceived as beneficial to Beidian intermediaries. We select 35 top posts with the most clicks from the section for analysis.  

Complementary to the forum analysis, we conduct an interview study on 15 randomly sampled Beidian intermediaries. The interviews are conducted in remote and in Mandarin, where we probe into what and how they are acting on Beidian and audio-tape the interviews after receiving their oral consent. The recordings are then manually transcribed by one Mandarin-speaking member of the research team.

To better protect the privacy of the members of the forum and the interviewees, we use pseudo-anonymous names and remove the identifiable information. Here names start with B (e.g., B1) represent members of the forum, where B3, B4, B5, B14, B23 and B24 are recognized as successful intermediaries by Beidian officially. Names start with I (e.g., I1) represent interviewees.

\section{What Do Intermediaries Do? }

\begin{figure} [t]
\centering
\subfigure[Distribution of fan number for intermediaries] {
\label{fig:FanNumDisAll}
\includegraphics[width=.49\textwidth]{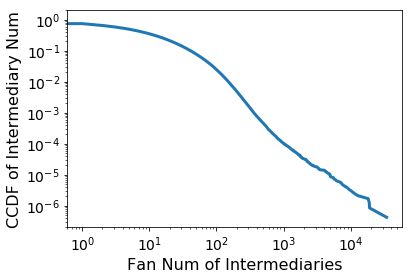}}
\subfigure[Distribution of total deal number for intermediaries] {
\label{fig:DealNumDisAll}
\includegraphics[width=.49\textwidth]{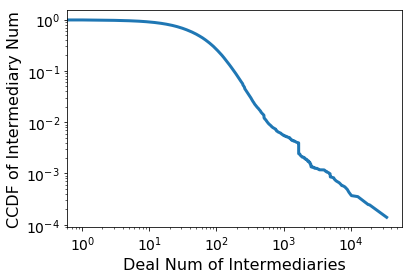}} \\

\caption{Basic characteristic distribution of intermediaries.} \label{fig:BasicDisAll}
\end{figure}

To uncover the characteristics of intermediaries on Beidian, we first show the basic distribution of intermediaries. Specifically, Fig. \ref{fig:FanNumDisAll} and Fig. \ref{fig:DealNumDisAll} delineate the number of fans and the number of purchases made under each intermediary, respectively. Here fans represent potential customers who have either registered or purchased through the intermediary's link. As shown in the figures, the vast majority of intermediaries do not have many fans, for example, only 2.5\% of all intermediaries have more than 100 fans. On the other hand, the number of deals made under each intermediary is relatively larger, for example, 26.3\% of all intermediaries have closed more than 100 orders. 

To eliminate the bias induced by inactive or dormant intermediaries, we rule out intermediaries who have invited less than 5 people or conducted less than 5 deals in total during our observation period, which takes up 50.9\% and 4.8\% of all intermediaries respectively. Note that the portion of intermediaries ruled out for inviting less than 5 people is relatively high due to several reasons: 1) some intermediaries exited the platform; 2) some intermediaries purchased the items required to qualify as intermediaries by the platform, yet did not intend to conduct sales as actual active intermediaries. We decide to set such thresholds for our analysis because including these cases won't add to our understanding of market intermediaries. To check the robustness of the threshold setting, we vary the threshold level and obtained similar results across different threshold settings.

On Beidian, intermediaries' sales are completed through sharing links and recommending products via social media applications, such as WeChat. To examine the effectiveness of intermediaries' recommendations, we investigate the relationship between intermediaries' recommendation proportion of an item and the proportion that the corresponding item takes up in purchases made through the intermediary. As demonstrated in Fig. \ref{fig:RecBuyOwnAll}, these two variables demonstrate a positive correlation, suggesting that intermediaries' recommendations are positively perceived and adopted by consumers.  
Items more likely to be consumed are exactly those items recommended more by intermediaries, which corroborates the effectiveness of intermediaries' dedication. 

To better understand the role of intermediaries, we delve into the strategies and forms they deploy to connect social e-commerce platforms and users based on their real-world close social contacts. In the following, we present their two salient roles: as local trend detectors and as socially-connected convenience stores. 

\subsection{Recommending Products as Local Trend Detectors}

\begin{figure} [t]
\centering
\subfigure[Total recommendations vs. total deal] {
\label{fig:RecBuyOwnAll}
\includegraphics[width=.32\textwidth]{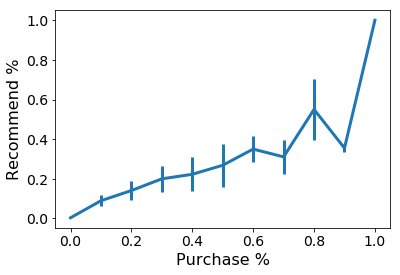}}
\subfigure[Next week's recommendations vs. last week's deal] {
\label{fig:RecBuyOwn7}
\includegraphics[width=.32\textwidth]{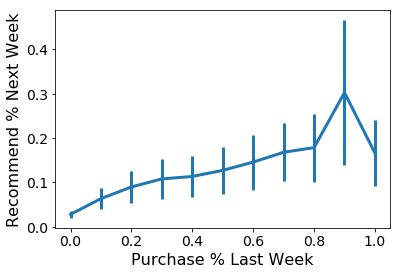}}
\subfigure[Next 3 days' recommendations vs. last 3 days' deal] {
\label{fig:RecBuyOwn3}
\includegraphics[width=.32\textwidth]{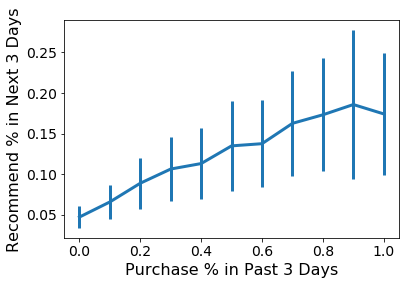}} 
\caption{Relationship between intermediaries' deal percentage and recommendation percentage.} \label{RecBuyOwn}
\end{figure}

To understand intermediaries' choices of recommendation in detail, we take into consideration the time dimension and examine whether consumers' historical purchases through an intermediary can influence an intermediary's recommendation choices. To reach the goal, we partition purchases and recommendations into time bins of weeks and three days, respectively. 
Fig. \ref{fig:RecBuyOwn7} depicts the relationship between intermediaries' last week's sales and the next week's recommendations, and the relationship between intermediaries' last 3 days' sales and the next 3 days' recommendations is delineated in Fig. \ref{fig:RecBuyOwn3}. Results show that past sales exert a positive influence on item promotion. To be more specific, the more purchases of an item are made through an intermediary in the past few days, the more likely that the intermediary will recommend the corresponding item to his or her potential consumers. Our qualitative study corroborates our quantitative analysis, as one intermediary explains:
\emph{"I usually recommend the items that sell well today ... they suit customers' needs ... If I decide to mainly promote this item today, I will promote it several times ... and provide screenshots of orders ... People have conformity. If he/she sees that others are buying, he/she may want to buy to try him/herself."} (B14)

Another intermediary underlines the importance of understanding consumers' needs: \emph{"the business of Beidian rests on WeChat friends and acquaintances ... you should choose products that meet the needs of your friends. It is better to recommend products according to your friends' characteristics or consumption ability."}(B1) Intermediaries' past sales indicate local communities' purchase propensity towards a particular item. Once the intermediary recommends that item to his or her potential clients, these people who share similar interests and needs regarding the item will be more likely to place an order. In turn, intermediaries also have a stronger willingness to promote and re-promote the item. The key mechanism working here is imitation, \emph{i.e.}, when acknowledging others in the community who have purchased an item, some users are likely to conform to others' purchases. As items with past sales are more likely to induce further purchases, intermediaries are more likely to recommend such items.

The discussion so far considers individual intermediaries who act as recommenders of products to consumers. Yet, intermediaries are also collectively organized into communities, such as WeChat group to exchange their experience and learn from best practices. As mentioned by our informants B13 and B15, they have specific WeChat groups organized only for intermediaries, where they interact, share selling tricks and exchange promotion experiences. These self-organized groups serve as "an informal college" for intermediaries to learn: \emph{"people see the information about what items are selling well in intermediary communities. As it turns out, these products are easier for you to sell because of their own transmissibility."} (B1) As a result, intermediaries follow the trend set by successful precedents and turn these products to their main sales for the day. Sometimes they not only refer to others' item choices, but borrow \emph{"others' texts and words for promotion posts directly, which can be used as references"} (B6) as well.

\begin{figure} [t]
\centering
\subfigure[Total recommendations vs. total deal] {
\label{fig:RecBuyComAll}
\includegraphics[width=.32\textwidth]{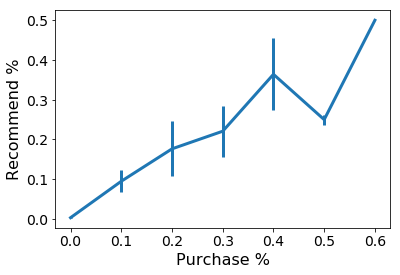}}
\subfigure[Next week's recommendations vs. last week's deal] {
\label{fig:RecBuyCom7}
\includegraphics[width=.32\textwidth]{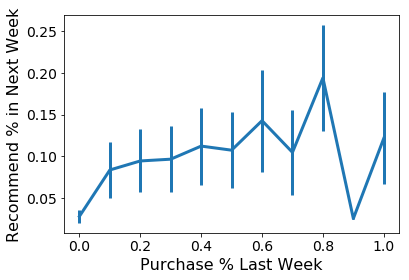}}
\subfigure[Next 3 days' recommendations vs. last 3 days' deal] {
\label{fig:RecBuyCom3}
\includegraphics[width=.32\textwidth]{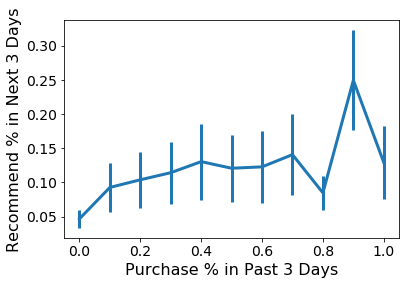}} 
\caption{Relationship between deal percentage in the community and recommendation percentage.} \label{RecBuyCom}
\end{figure}

To empirically examine whether and how an intermediary can be impacted by other intermediaries, especially those intermediaries in direct contacts, we investigate the relationship between consumers' past purchases through intermediaries in his or her community and the intermediaries' choices of recommendation. Here a community refers to people who are more likely to be socially acquainted with each other or more likely of sharing the same characteristics. More specifically, we define a community of intermediaries as those who are invited by the same upstream intermediary to join Beidian in the first place. The assumption here is that these intermediaries are either connected in their real lives or they share some similarities in some aspects. 
Fig. \ref{fig:RecBuyComAll} demonstrates the relationship between intermediaries' proportion of recommendation and the proportion of consumer purchases through intermediaries in their communities. It suggests that intermediaries' probability of recommending a particular item positively correlates with communities' sales volume of the same item.

However, the figure cannot confirm that closely connected intermediaries can impose an impact on each other, because it is likely that recommendations of a particular item can be made simultaneously in a community. To address this problem, we turn to the sequential analysis of intermediaries' recommendations and sales completed with their communities. Similar to the aforementioned analysis, we partition purchases and recommendations into time bins of weeks and three days, respectively. Fig.\ref{fig:RecBuyCom7} shows the relationship between communities' last week's sales and intermediaries' next week's recommendations. Fig. \ref{fig:RecBuyCom3} shows the relationship between communities' past 3 days' sales and intermediaries' next 3 days' recommendations. Patterns in both figures suggest that a positive correlation exists between the percentage that an item takes up in intermediaries' future recommendation and the percentage it takes up in the communities' past sales (note that in Fig. \ref{fig:RecBuyCom7}, only two records share a community sales percentage of around 0.9 in the past week and therefore induce bias).

To sum up, our analysis demonstrates that intermediaries utilize both their own past sales and community members' past sales to make recommendations. If we regard the percentage of past sales as a sign of the local trend, \emph{i.e.}, the popularity of the item locally, these intermediaries act as local trend detectors: sorting and collecting information of potential consumers in their surroundings, analyzing items that are more likely to be led to deals, and consequently recommending and promoting them.

\subsection{Serving Customers as Socially-connected Convenience Stores}

\begin{figure} [t]
\centering
\subfigure[Customer number of intermediaries] {\label{fig:FanNumDis}
\includegraphics[width=.49\textwidth]{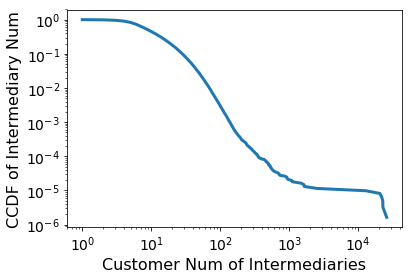}}
\subfigure[Deal number of consumer-intermediary pairs] {\label{fig:DealNumDis}
\includegraphics[width=.49\textwidth]{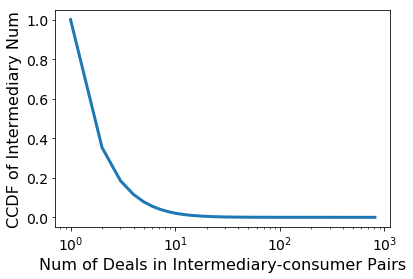}} \\
\subfigure[Sales volumes of consumer-intermediary pairs] {\label{fig:SalesVolDis}
\includegraphics[width=.49\textwidth]{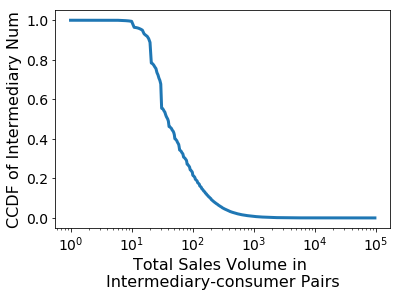}} 
\subfigure[Average price of deals within consumer-intermediary pairs ]{\label{fig:PriceDis}
\includegraphics[width=.49\textwidth]{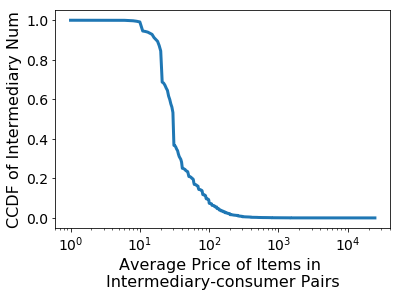}}
\caption{Distribution of purchase-relating features of intermediaries.} \label{BasicDis}
\end{figure}

After presenting how intermediaries make recommendation choices, we now turn to explain how they serve their consumers. First, we present the distribution of customer number across intermediaries. As depicted in Fig.~\ref{fig:FanNumDis}, the vast majority of intermediaries do not have a large number of customers. To be more specific, 55.0\% of the intermediaries have fewer than 10 customers, and the number of customers for 99.7\% intermediaries' does not exceed 100. This suggests that intermediaries on social e-commerce platforms are similar to convenience stores that own a limited number of purchasers~\cite{zeng2012retailing}.

\begin{figure} [t]
\centering
\subfigure{\includegraphics[width=.49\textwidth]{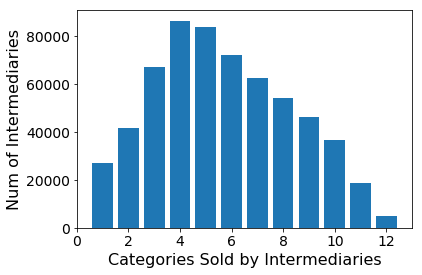}}
\caption{Distribution of product category count sold by intermediaries.} \label{CatNum}
\end{figure}

In terms of interactions between consumers and intermediaries, we turn to the number of deals, total sales volume and the average item price per sale, respectively. Fig.~\ref{fig:DealNumDis} demonstrates the distribution of the number of deals completed between a consumer and an intermediary. As shown by the picture, 35\% of the consumer-intermediary pairs have more than two records of purchases during the recorded interval of 9 months, but only 2.1\% have consumed 10 times or more. This is in accordance with convenience stores' characteristic of instant consumption~\cite{zeng2012retailing}, where most customers purchase purposefully~\cite{china2004retail} and therefore not a great number of deals are made per consumer. As for total sales volume, as depicted in Fig.~\ref{fig:SalesVolDis}, the vast majority of consumers' total value of purchases from an intermediary lies within the interval between 10 RMB yuan and 1000 RMB yuan. The portion between 10 and 100 RMB yuan takes up 73.5\% of all consumer-intermediary pairs, and that between 100 RMB yuan and 1000 RMB yuan stands for 23.3\%. This is consistent with the small scale of convenience stores~\cite{levy1996essentials}. When we consider the sales volume per purchase, not much amount is consumed per deal within most consumer-intermediary pairs (see Fig.~\ref{fig:PriceDis}). Specifically, 91.6\% of the deals lie within the interval of 10 RMB yuan and 100 RMB yuan. This is consistent with convenience stores' property that those not-so-costly daily necessities make the majority of purchases. Results concerning the number of categories sold by intermediaries further corroborate our labeling of Beidian intermediaries as convenience stores. As demonstrated in Fig.~\ref{CatNum}, the vast majority of intermediaries have successfully promoted a wide range of items covering different categories rather than focusing on a particular or an exclusive list of categories.

Our qualitative study further supports our quantitative findings. In terms of customer size, many intermediaries' customers are rather limited, sometimes confined to "the familiar" (I1) or "friends around" (I4), or perhaps only 5 relatives and friends (I2). In terms of deal number, intermediaries admit that many users do not place many deals from them. As mentioned by B11, \emph{"many people just do not move along after placing one or two orders"}. But, on the other hand, the convenience store like nature also has its advantages. As an informant explains, \emph{"it is just like we shop in the grocery store ... As long as the grocery store is there, you will always come by to see and buy something. You are always there, when others need you. Trust is then established and everything will be easy."}(B11) Some other intermediaries note that \emph{"there is no need to think that every group should be 100 percent active. It is impossible and unnecessary. Only a small number of people can contribute to most value for you in the end."}(B9) When describing the items they recommend, they mention mostly daily items like "mango, lemon, duck eggs, rice dumplings, towel, laundry detergent" (B8) which "do not cost much for trial" (B14). Taking these features together, we are confident in labeling intermediaries as convenience stores. While the audience for buying is relatively limited, and the frequency people make purchases is not very high, the products purchased are from a relatively wide range, are not so costly and in most cases are daily necessities.

\begin{figure} [t]
\centering
\subfigure[City similarity of intermediaries and their consumers] {\label{CitySim}\includegraphics[width=.49\textwidth]{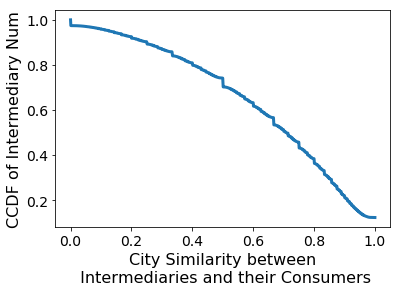}}
\subfigure[Province similarity of intermediaries and their consumers]{\label{ProSim}\includegraphics[width=.49\textwidth]{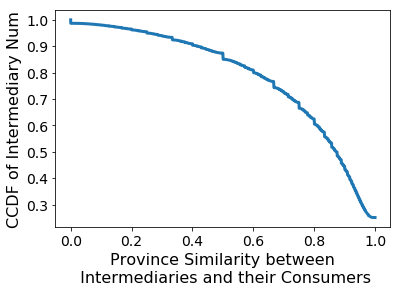}} 
\caption{Distribution of geographical similarity of intermediaries and their consumers.}
\label{BasicGeo}
\end{figure}

Convenience stores in its conventional sense are usually of a small scale and are easy to reach within 5 minutes' walk in a neighborhood~\cite{china2004retail}.  However, when we examine geographic features of social e-commerce convenience stores, counter-intuitive findings surface. As illustrated in Fig.~\ref{BasicGeo}, market intermediaries' outreach via social media applications goes beyond geographical locations. Here we define the probability that a consumer of an intermediary locates in the same place as the intermediary as \emph{similarity} of one intermediary with his or her consumers. From Fig.~\ref{CitySim}, we observe that only 12.3\% of the intermediaries are similar to all their customers on the city level. On a provincial level, as shown in Fig.~\ref{ProSim}, that portion raises to 25.1\%, which still makes only a small part of all intermediaries. Therefore, we suggest that a strong locality is not the case for most intermediaries' promotion. The majority of intermediaries have managed to extend their sales to a larger geographic scale, which is not the case for convenience stores. Distant places are thus connected and long-distance retailing is enabled through the combination of real-world social relations and e-commerce platforms. 

Our qualitative study provides more details on the surprising findings of the geographical reach beyond local. When speaking of the constitution of the customers in their sales groups, intermediaries frequently mention socially-closed ones and provide reasonable explanations: \emph{"friends, relatives, classmates, colleagues, and the closely connected ... This group of people are more supportive ... trust makes it easier to place orders ... Then the secondary network, when we call on friends and relatives who join the group to help pull people they know into the group (also referred to as friends' friends by some intermediaries). These people are also more likely to place orders because there are trusted people in the middle to act as bridges ... Sometimes also strangers with great potential to buy ... but 95\% of the strangers lurk in groups."} (B12)

Similar accounts are often mentioned by intermediaries both from the Beidian forum and in the interviews, and some intermediaries even only recommend to the familiar ones: \emph{"None of my fans are strangers ... Strangers are defensive. They don't take these things from you."}(I1) With trust rooted in close social relationships and better knowledge of what would be interesting to strong ties, intermediaries identify the socially closed ones as their top priorities for audience selection. As a result, in intermediaries' sales, the geographical closeness of convenience stores is turned to social closeness.

In summary, most intermediaries' promotion on Beidian takes a form that resembles convenience stores. However, rather than exposing to physically-close customers, intermediaries' sales target socially-related ones. This leads to a new form of "socially-connected convenience stores", where intermediaries promote items in the same way as convenience stores, but to potential consumers socially rather than geographically close to them.
\section{Why do some Intermediaries stand out?}
To better understand intermediaries, in this section we focus on the heterogeneity among intermediaries and investigate why some intermediaries are more effective local trend detectors and socially-connected convenience stores than others. More specifically, we examine how 1) recommendation strategies and 2) transaction modes lead to the differences. Capturing what makes some intermediaries stand out in the crowd in a more detailed manner will contribute to our knowledge of the market dynamics of social e-commerce. 

\subsection{Recommendation Strategies}
Despite the overall picture of intermediaries we demonstrate, we are aware that intermediaries may differ in their behaviors and strategies in connecting platforms and users. To unpack the heterogeneity, we cluster them into groups based on their recommendation behaviors and correlate the clusters with their performance as intermediaries.

\subsubsection{Classification of Intermediaries}

\begin{figure*}[h]
  \centering
      \begin{subfigure}{
        \includegraphics[width=0.95\textwidth]{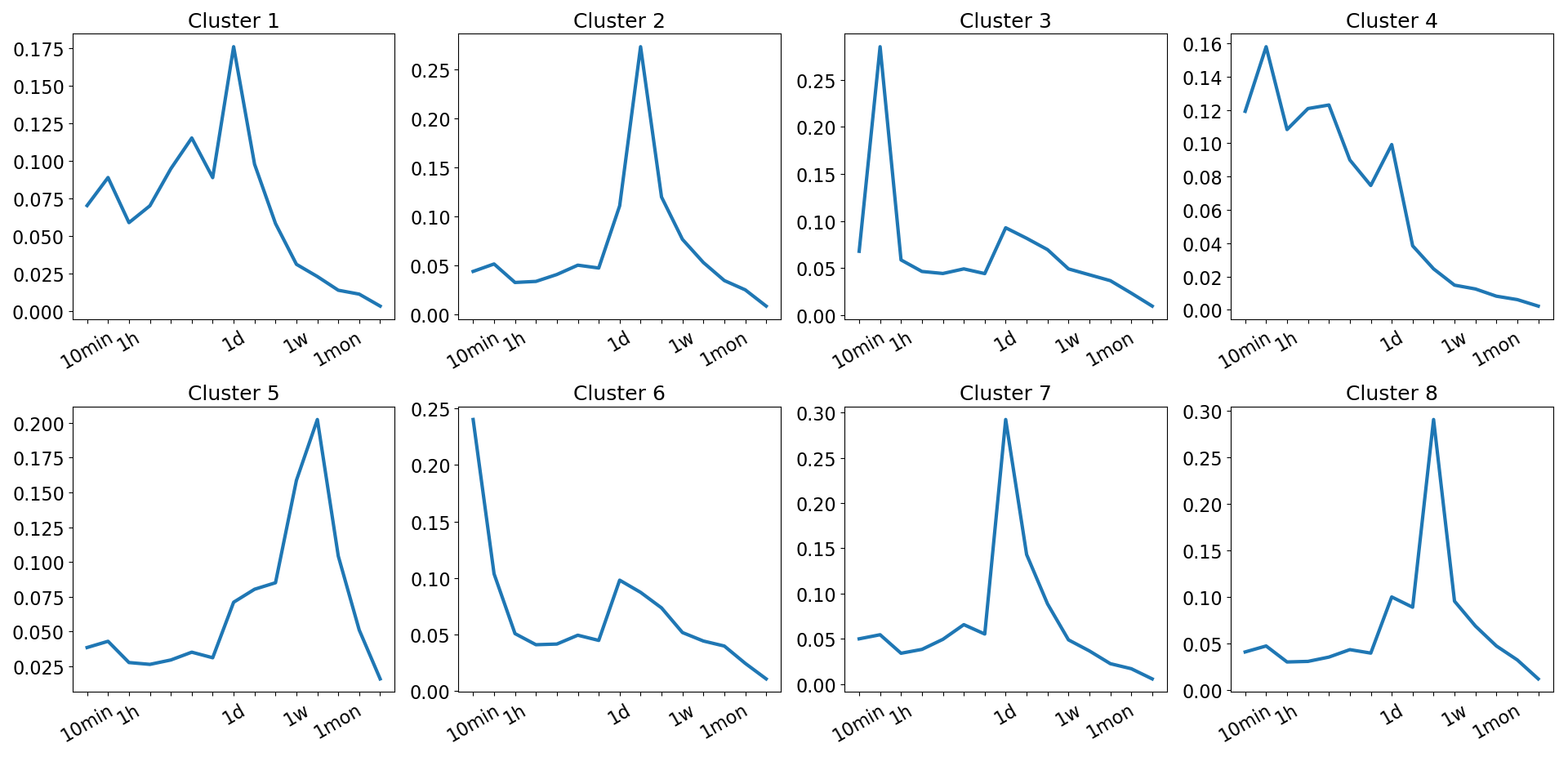}}
      \end{subfigure}
      \caption{Centroid re-recommendation curves for 8 identified clusters.}
  \label{ClusterCenter}
\end{figure*}

To mine intermediaries' recommendation patterns, we follow the techniques of revisitation curve method utilized in \cite{adar2008large,jones2015revisitation,cao2018revisitation} to analyze intermediaries' re-recommendation behaviors. The re-recommendation curves depict the number of times (normalized) an intermediary is likely to re-conduct the behavior of recommendation within a predefined time interval. Following \cite{adar2008large,jones2015revisitation,cao2018revisitation}, we use an exponential scale to construct bins for our re-recommendation curves. To avoid the possibility that too frequent records are generated from one action of recommendation and several consecutive recommendations are actually one action of sharing to different people, we rule out those re-recommendation record with intervals shorter than 5 minutes, and time bins of < 10 minutes, 30 minutes, 1 hour, 2 hours, 4 hours, 8 hours, 12 hours, 1 day, 2 days, 4 days, 1 week, 2 weeks, 1 month, 2 months, >2 months are used. Following \cite{jones2015revisitation} and \cite{cao2018revisitation}, we apply the K-means algorithm to cluster intermediaries, leveraging "Euclidean distance" as measurements for distances on multiple $k$ values. To decide the optimal $k$, we construct an elbow plot utilizing the sum of square errors from samples to identified cluster centroids with $k$ ranging from 2 to 20. We finally choose $k=8$, the elbow point where the sum of square errors does not drop significantly. The centroids of the identified clusters are demonstrated in Fig. \ref{ClusterCenter}. 

\subsubsection{Recommendation Behaviors of Successful Intermediaries}
\begin{figure} [t]
\centering
\subfigure[Average number of consumers for intermediaries in different clusters] {
\label{fig:ClusterBuyer}
\includegraphics[width=.32\textwidth]{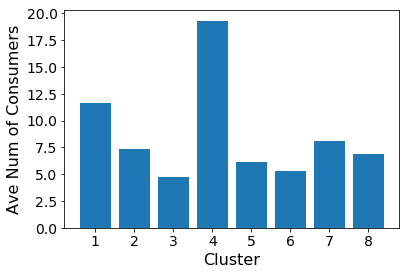}}
\subfigure[Average number of deals made for intermediaries in different clusters] {
\label{fig:ClusterDeal}
\includegraphics[width=.32\textwidth]{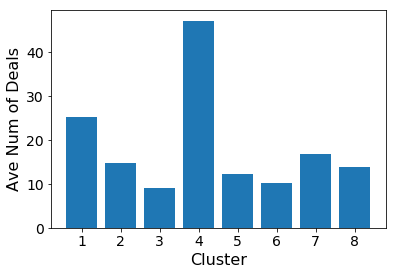}}
\subfigure[Average total sales volume for intermediaries in different clusters] {
\label{fig:ClusterMoney}
\includegraphics[width=.32\textwidth]{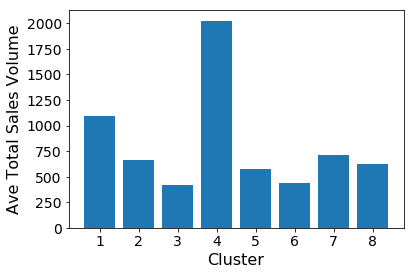}}\\
\subfigure[Average amount per deal for intermediaries in different clusters] {
\label{fig:ClusterMoneyPerDeal}
\includegraphics[width=.32\textwidth]{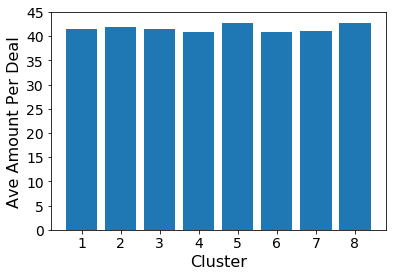}}
\subfigure[Average number of deals for intermediary-consumer pairs in different clusters] {
\label{fig:ClusterPerDeal}
\includegraphics[width=.32\textwidth]{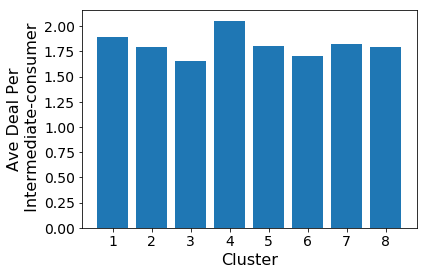}}
\subfigure[Average total sales volume for intermediary-consumer pairs in different clusters] {
\label{fig:ClusterPerMoney}
\includegraphics[width=.32\textwidth]{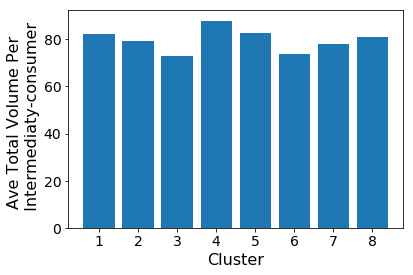}}
\caption{Purchase-relating features of intermediaries in different clusters.} \label{ClusterFea}
\end{figure}

To determine how well these intermediaries are taking their roles, we calculate the average consumer numbers, the average sales numbers, and the average total sales volumes for intermediaries in different clusters. As demonstrated in Fig. \ref{fig:ClusterBuyer}, Fig. \ref{fig:ClusterDeal} and Fig. \ref{fig:ClusterMoney}, intermediaries in cluster 4 not only reached the largest number of people on average, but also completed the largest number of deals and made the highest turnover across all clusters. We regard intermediaries in this cluster as "successful intermediaries" and we here delve into their strategies more closely.

In terms of deal number per intermediary-consumer pair, as illustrated in Fig. \ref{fig:ClusterMoneyPerDeal}, the average number of deals among intermediary-consumer pairs is higher than intermediaries recognized as within other clusters. In terms of the price of the items to be consumed, the average sales per unit in this cluster does not show much difference from intermediaries in other clusters, if not slightly lower than some clusters (see Fig. \ref{fig:ClusterPerDeal}). When combining these two factors together and investigating the average total sales volume within intermediary-consumer pairs, intermediaries in cluster 4 managed to persuade their customers to consume more on average (see Fig. \ref{fig:ClusterPerMoney}). Taking all these features together, our analysis suggests that these successful intermediaries do not necessarily aim at completing big deals. Casting a wider net by increasing the number of deals with consumers contributes to their success relative to other clusters. 

For further understanding of recommendation behaviors, we turn to shapes of the re-recommendation curves of the 8 identified clusters. As demonstrated in Fig. \ref{ClusterCenter}, the peak of cluster 5's and cluster 8's re-recommendation curves is around a week. Re-recommendation peaks of cluster 1, cluster 2 and cluster 7 are on a daily basis. Cluster 4 reaches their climaxes around 30 minutes and its performance resembles the shape of a heavy tail. Cluster 3 shares the same peak as cluster 4, but concentrates on a short period. Intermediaries in cluster 6 are most likely to re-recommend within 10 minutes and their actions of re-recommending concentrate within 1 hour.

Correlating the shapes of clusters' centroid re-recommendation curves and their corresponding sales performances shows that clusters that peak in a not-long re-recommendation interval with heavy tails achieve better sales. This is consistent with the theory of bursty behavior~\cite{barabasi2005origin}. Intermediaries' behaviors of recommendation can be seen as a decision-based queuing process. Intermediaries expose to items randomly, and it is up to intermediaries to decide what items to prioritize for recommendation. Utilizing the strategy of recommending items through a priority list and always recommending the more preferred, a long-tail distribution is likely to be induced. Therefore, the strategy to be proposed is to establish a recommendation priority list. In the absence of best sellers, recommendations are made infrequently. However, when best sellers come, these successful intermediaries recommend it recursively every time a not-long period passes.

Our qualitative evidence further corroborates our suggestion. Because of the privacy-preserving limitations of the data, we are not able to directly link the respondents in qualitative studies to their IDs in quantitative studies. However, as mentioned in Section 3.3, some intermediaries are certified as successful by Beidian officially. We here use them to represent successful intermediaries for analysis. 

As mentioned by successful intermediaries, more frequent recommendations don't necessarily lead to better sales. B3 explained, \emph{"always recommending will only get customers to quit"}, leading potential customers to treat the recommendations as spams and block the intermediary. Therefore, "fewer but more refined recommendations" (B3) are preferred and in most cases these successful intermediaries intentionally \emph{"control the total amount of recommendations per day"} (B23). Their rather focused recommendations mostly relate to the "must-have specials" (B3) or "hot vogues" (B23). Whenever a "top seller" (B23) come, they \emph{"spare no effort to promote that one"} (B23), \emph{"place the item in the notice of the group and send it repeatedly every once in a while ... with pictures and shopping list"} (B23) and \emph{"post the screenshots of the items for times"} (B14). By doing so, successful intermediates maintain an equilibrium between tension and relaxation. They usually do not disturb potential customers much; when top sellers come, their focused dedication can greatly facilitate transactions. As a result, sustainable sale environments are established.

\subsection{Transaction Modes}
To obtain a deeper comprehension of the identified successful intermediaries, we further examine their transaction modes and contrast with the overall patterns of all intermediaries.

\begin{figure} [t]
\centering
\subfigure[Customer number of intermediaries] {\label{BuyNum4}\includegraphics[width=.49\textwidth]{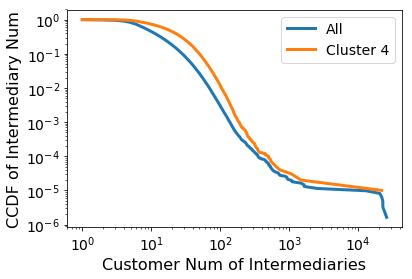}}
\subfigure[Deal number of consumer-intermediary pairs] {\label{DealNum4}\includegraphics[width=.49\textwidth]{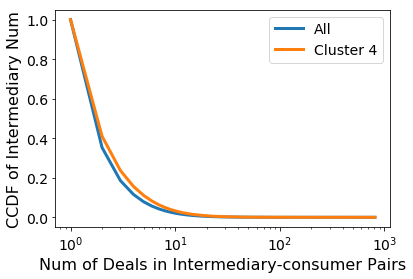}} \\
\subfigure[Sales volumes of consumer-intermediary pairs] {\label{Money4}\includegraphics[width=.49\textwidth]{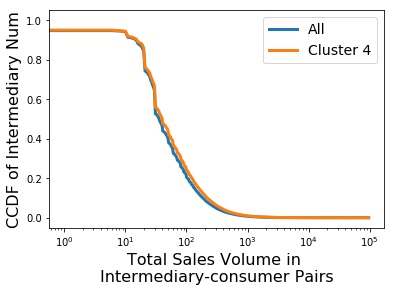}} 
\subfigure[Average prices of purchases within consumer-intermediary pairs ]{\label{PerMoney4}\includegraphics[width=.49\textwidth]{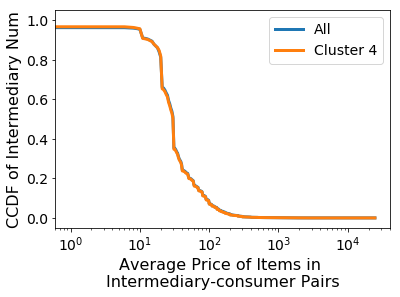}}
\caption{Distribution of purchase-relating features of intermediaries in cluster 4 vs all.} \label{BasicDis4}
\end{figure}

In terms of customer number, as shown in Fig. \ref{BuyNum4}, intermediaries in cluster 4 have a broader customer base (Kolmogorov–Smirnov test$=0.303$, $p<0.001$) compared to the overall case. We deem that as long as more customers are exposed to the intermediaries, more purchases are likely to be generated and therefore the intermediaries would become more successful in promoting their commodities. In addition, it is worth mentioning that the portion of intermediaries with an extremely large number of consumers in cluster 4 is not as large as the overall. This could be explained by the limitation of intermediaries' capacity. A single intermediary does not have the ability to cope with too many customers, and therefore, it is likely that these successful intermediaries do not try to broaden their own consumer bases once their businesses achieve a certain scale. In terms of deal number within consumer-intermediary pairs, we show in Fig.~\ref{DealNum4} that significantly more deals are made per pair in cluster 4 than overall, but the difference shown is relatively smaller (Kolmogorov–Smirnov test$=0.057$, $p<0.001$). It is believed that more deals per consumer would contribute to more sales under the circumstance of the same number of consumers. Speaking of total sales volume, similar cases are also spotted (see Fig.~\ref{Money4}), where the total sales volume of intermediaries in cluster 4 is significantly higher than the overall (Kolmogorov–Smirnov test$=0.039$, $p<0.001$). We further depict the distribution of the average price of the items recommended by intermediaries in cluster 4 vs. all intermediaries in Fig.~\ref{PerMoney4}. Kolmogorov–Smirnov test shows that the difference between the distribution of cluster 4's item price and the overall case is significant, but the difference may not be that much (Kolmogorov–Smirnov test$=0.001$, $p<0.001$). This probably indicates that although the price of the items can contribute to intermediaries' success of recommendations, the contribution may be rather limited in size.

\begin{figure} [t]
\centering
\subfigure{\includegraphics[width=.49\textwidth]{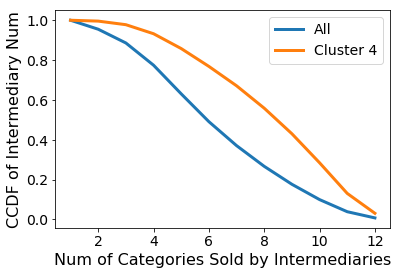}}
\caption{Distribution of number of item categories sold by intermediaries in cluster 4 vs all.} \label{CatNum4}
\end{figure}

\begin{figure} [t]
\centering
\subfigure{\includegraphics[width=.49\textwidth]{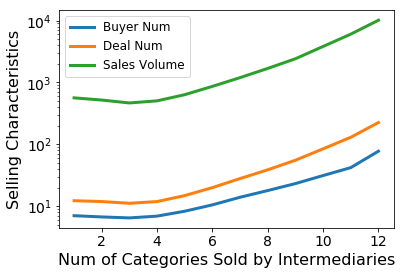}}
\caption{Average number of buyers, deals, total sales volume in terms of item categories sold by intermediaries.} \label{CatNum4Value}
\end{figure}

When we turn to the diversity of intermediaries' sales, we show in Fig.~\ref{CatNum4} a significant difference between cluster 4 and the overall picture (Kolmogorov–Smirnov test$=0.300$, $p<0.001$). This figure suggests that the identified successful intermediaries in cluster 4 have a higher tendency of promoting products under the realm of more categories. To further corroborate our findings, we investigate the relationship between the number of categories sold by intermediaries and average buyer number, deal number and total sales volume of intermediaries, respectively. As demonstrated in Fig.~\ref{CatNum4Value}, these three factors positively correlate with selling categories, which indicates that to become better intermediaries, it is recommended that one should broaden his or her range of commodities as much as possible.

\begin{figure} [t]
\centering
\subfigure[City similarity of intermediaries and their consumers] {\label{CityGeo4}\includegraphics[width=.49\textwidth]{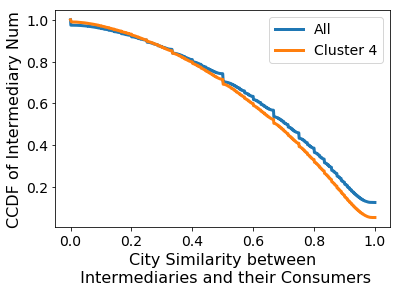}}
\subfigure[Province similarity of intermediaries and their consumers]{\label{ProGeo4}\includegraphics[width=.49\textwidth]{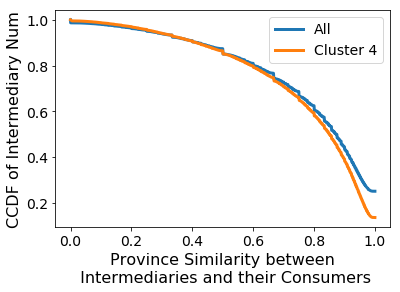}} 
\caption{Distribution of geographical similarity of intermediaries and their consumers in cluster 4 vs all.}
\label{BasicGeo4}
\end{figure}

For geographic features, we plot the distribution of city similarity and province similarity between intermediaries and their corresponding consumers in cluster 4 and overall in Fig.~\ref{CityGeo4} (Kolmogorov–Smirnov test$=0.072$, $p<0.001$) and Fig.~\ref{ProGeo4} (Kolmogorov–Smirnov test$=0.115$, $p<0.001$), respectively. It can be drawn from the figures that successful intermediaries' consumers are less similar to them geographically, which indicates that these successful intermediaries in cluster 4 are better at boosting sales beyond their geographical proximity.  

These findings are validated by our qualitative study. Successful intermediaries share a wider range of customers and can better retain customers through the utilization of "attractive product descriptions" (B3, B4, B14, B23) and better management of sales groups to "make the groups active" (B3, B5, B24), although the products recommended may be ordinary items like pillows or tableware (B3), mangoes or cakes (B14), etc. This is partly due to their believes that \emph{"adopting the mode of low unit price, small profits but quick turnovers is beneficial"} (B3). In terms of the kinds of items they sell, some successful intermediaries name their sales groups as grocery stores directly (B23), recommending various products such as "noodles, baby picture books, hangers, towels and hair conditioners" (B23) all together at once. What's more, they regard it important to \emph{"keep up with the recent trend and recommend the appropriate ones accordingly"} (B3), and thus constantly bring new items to be recommended to prevent customers from being bored. As for relationships with customers, they treasure more on the social closeness within sales groups rather than geographical similarities (B24) because these people are more likely to share analogous needs in terms of purchases.

Gathering the aforementioned characteristics of successful intermediaries, we propose that more effective intermediaries are those who hold a wide range of categories of items and attract customers to their capacity. Being a small-scale generalist makes a more successful "socially-connected convenience stores". In addition, they actively mobilize geographically-distant but socially-close relationships and extend their scale of intermediaries to a moderate scale.
\section{Discussion}

Drawing on large-scale quantitative studies and in-depth qualitative analysis, we systematically analyze the role of intermediaries on the social e-commerce platform Beidian. By fleshing out the characteristics, recommendation strategies and transaction modes of these intermediaries, our study sheds light on the expert middlemen who are critical for the marketplace dynamics of social e-commerce platforms, where novel insights and potential benefits can be obtained.

\subsection{Manual Recommender Systems}
Through the introduction of intermediaries, social e-commerce enjoys the welfare of manual recommender systems. Intermediaries greatly shape who becomes consumers and decide what to recommend to potential consumers on social e-commerce platforms. Unlike the widely-adopted mechanical recommender systems where machines determine the whole process of recommending commodities basing on past behaviors and proclivities~\cite{aggarwal2016recommender}, social e-commerce relies on intermediaries to enact the action of recommendation and formulate strategies with regard to what to recommend, how to recommend and when to recommend. In this way, the recommendation is transferred from machines to humans.

With social relationships as the infrastructure, intermediaries have more background information of the potential customers and a better understanding of their interests and needs, cultivating the effectiveness of manual recommender systems. Therefore, they are more likely to promote suitable items to appropriate people, resulting in better sales. As illustrated in Section 4.1, intermediaries act as local trend detectors on social e-commerce platforms, where they grasp what items would be suitable for recommending to a certain group of people. By doing so, human knowledge is integrated into the recommendation decision procedure, where intermediaries subjectively judge what are popular and can continue to sell well within their social circle of potential consumers. 

Another advantage of intermediaries as manual recommender systems is proactively pushing commodities directly to potential consumers via social media applications. In traditional e-commerce, only after entering the e-commerce platforms can users be exposed to recommendations. However, in social e-commerce, once an intermediary discovers a potential best seller suitable for his or her social cycle, he or she can more directly expose consumers to the item through instant messaging services such as WeChat messages and WeChat moments. Some e-commerce platforms nowadays push recommendation information regularly through their applications. However, they suffer from 1) not being so timely and not matching consumers' tastes or preferences, and 2) resembling spams. With the utilization of intermediaries as manual recommender systems, both limitations can be addressed to some degree. For intermediaries in social e-commerce, existing social ties are used for promotions. Provided that one is more acquainted with an intermediary, he or she is less likely to become frustrated by the intermediaries' recommendations and can even possibly generate unexpected purchases because of the social cost of rejection and benefits of social trust, as noted by~\cite{cheng2018diffusion}. Therefore, these recommendations are less likely to be regarded as spams, which enhances the actual efficacy of recommendations.

\subsection{Social Grocers}
As shown by aforementioned quantitative and qualitative analysis, most intermediaries promote commodities in a way similar to convenience stores, where 1) the number of customers is relatively limited, 2) the items purchased belong to daily necessities, 3) satisfying people's unique needs is of great importance and 4) a medium variety of products are recommended. However, different from the traditional grocery stores, intermediaries' sales reach those socially-connected beyond geographical constraints. The findings that successful intermediaries perform better in these aspects further strengthen our confidence in our results. We provide two possible interpretations here. On the one hand, these intermediaries who are socially connected with the targeting consumers have a higher chance of acknowledging consumers' comprehensive needs. Therefore, there is a wider range of suitable items for them to recommend to target consumers, which results in higher chances of purchases. On the other hand, appropriate use of existing social ties between intermediaries and consumers is essential to successful sales, where too frequently bothering the same range of people can lead to abomination. Therefore, a moderate amount and frequency for recommendations would be important for the success of intermediaries, where a promotion strategy resembling convenience stores would be an applicable option.

\subsection{Implications}
Our study provides novel implications for practices in related areas. Firstly, our work demonstrates a means to incorporate humans into the loop of recommendation decision procedures. We suggest that when designing platforms of this kind, practitioners would benefit from taking more social factors into consideration and making better use of human efforts and relationships. Secondly, we provide an account showing the benefit of crowdsourcing and collective intelligence. From the perspective of platforms, in addition to recruiting Key Opinion Leaders (KOL) for promotions, social e-commerce shows the potential of enabling everyone to become promoters through the introduction of intermediaries. With appropriate incentive schema, intermediaries self-motivate themselves to promote items and can contribute to the competitive advantage of platforms. As we have shown, although most sales of single intermediaries may not be expanded to a large scale, aggregating their contribution can develop an enormous trading ecology for platforms. For the future design of social computing systems, it would be interesting and impactful to follow similar strategies: self-motivate individuals, connect individuals through technology-enabled interfaces (e.g., social media platforms), and achieve collective goals through well-designed platform mechanism.

\subsection{Limitations and Future Work}
There are some limitations in our work, which provide exciting avenues for future research. Firstly, our paper relies on the data from one single social e-commerce platform in China, and there is the potential of context bias for our findings. Future work can consider cross-platform analysis to enhance the generalization of our results. Secondly, our work has been exploratory in nature and revealed some characteristics of intermediaries' roles. Future studies can proceed on deeper analysis, such as using network analytic methods to identify influential intermediaries, investigate how recommendations spread through the intermediary network, and how influential intermediaries shape the ecology, or studying the language of intermediaries to better understand their persuasion strategies. Thirdly, due to the limitation of our data acquisition, the user behavior dataset only covers 9 months' records. We wish to see further studies extending our study to a longer period, which would enable analysis on the evolution of intermediaries and longitudinal tracking of intermediaries' behavioral changes.

\section{Conclusion}
This paper presents the first analysis on the role of intermediaries in social e-commerce through the combination of quantitative explorations and qualitative analysis. We highlight characteristics of intermediaries, identify successful intermediaries and analyze factors contributing to becoming successful intermediaries. We demonstrate that they take the role of 1) detecting and promoting local trends; and 2) selling in a convenience-store manner to socially-connected consumers. We show that to become more successful intermediaries, one should actively devote to promoting best sellers and broaden their sales in a social grocer manner. Our work points to the interesting nature of behaviors and sales on this new form of social e-commerce, which contributes to a better understanding of computer-supported cooperative work and social computing based marketing platforms.

\begin{acks}
This work was supported in part by The National Key Research and Development Program of China under grant 2018YFB1800804, the National Nature Science Foundation of China under U1936217, 61971267, 61972223, 61941117, 61861136003, Beijing Natural Science Foundation under L182038, Beijing National Research Center for Information Science and Technology under 20031887521, and research fund of Tsinghua University - Tencent Joint Laboratory for Internet Innovation Technology.
\end{acks}

\bibliographystyle{ACM-Reference-Format}
\bibliography{refer,other_refer}


\end{document}